\documentclass[
aps,prl, twocolumn,superscriptaddress, amsmath,amssymb
]{revtex4-1}

\usepackage{graphicx}
\usepackage{bm}
\newcommand{\ket}{\rangle}

\begin{document}

\title{$\Lambda$-Enhanced Imaging of Molecules in an Optical Trap}

\author{Lawrence W. Cheuk}
\email{lcheuk@g.harvard.edu} 
\author{Lo\"ic Anderegg}
\author{Benjamin L. Augenbraun}
\author{Yicheng Bao}
\author{Sean Burchesky}

\affiliation{Department of Physics, Harvard University, Cambridge, MA 02138, USA}
\affiliation{Harvard-MIT Center for Ultracold Atoms, Cambridge, MA 02138, USA}

\author{Wolfgang Ketterle}
\affiliation{Harvard-MIT Center for Ultracold Atoms, Cambridge, MA 02138, USA}
\affiliation{Department of Physics, Massachusetts Institute of Technology, Cambridge, MA 02139, USA }

\author{John M. Doyle} 
\affiliation{Department of Physics, Harvard University, Cambridge, MA 02138, USA}
\affiliation{Harvard-MIT Center for Ultracold Atoms, Cambridge, MA 02138, USA}

\date{\today}

\begin{abstract}
We report non-destructive imaging of optically trapped calcium monofluoride (CaF) molecules using in-situ $\Lambda$-enhanced gray molasses cooling. $200$ times more fluorescence is obtained compared to destructive on-resonance imaging, and the trapped molecules remain at a temperature of $20\,\mu\text{K}$. The achieved number of scattered photons makes possible non-destructive single-shot detection of single molecules with high fidelity. 

\end{abstract}

\maketitle

Ultracold molecules hold promise for many important applications, ranging from quantum simulation~\cite{zoller06,carr09,pupillo08,zoller07} and quantum information processing~\cite{demille02qi,yelin06,blackmore2018tweezerQI} to precision tests of fundamental physics~\cite{ACME14,hinds12,Lim2017,Kozyryev2017,carr09}. Recently, direct laser cooling of molecules has seen rapid progress. Starting from the first demonstrations of magneto-optical traps (MOTs)~\cite{barry14,norrgard16RF,steinecker16,Truppe2017subdoppler,anderegg17}, laser cooling to sub-Doppler temperatures, magnetic trapping and optical trapping of directly cooled molecules have all been achieved~\cite{Williams2018magtrap,McCarron2017magtrap,Anderegg2018ODTcooling}. 

Applications in quantum simulation and information processing demand high-fidelity detection of the molecules, which has been a focus of recent work~\cite{zeppendfeld2017detection}. Other applications, including precision measurement, can also benefit from improved detection. Typically, fluorescence imaging of trapped ultracold samples is destructive due to recoil heating from photon scattering. In recent years, advanced imaging techniques for atoms have circumvented such heating, achieving sensitivities to detect single atoms detection sensitivity. This has enabled quantum gas microscopy~\cite{Bakr2009microscope,Sherson2010mott,haller2015microscope,Cheuk2015microscope,Parsons2015microscope}, which has provided unprecedented microscopic access into quantum many-body systems. Furthermore, non-destructive imaging has opened up new routes to prepare quantum states, as has been demonstrated recently in optical tweezer experiments~\cite{lukin16array,barredo16array,Bernien201751atom}.

In this letter, we report on non-destructive imaging of optically trapped CaF molecules. We are able to scatter $2700$ photons while keeping $90\%$ of the molecules trapped at a temperature of $20\,\mu\text{K}$. Compared to standard on-resonance imaging, 200 times more photons are collected. At the heart of our imaging method is a cooling technique known, in the context of alkali atoms, as $\Lambda$-enhanced gray molasses cooling. Despite a more complex internal structure in CaF, we have identified a scheme wherein $\Lambda$-enhanced cooling can be implemented, and have used it to cool molecules in free-space to $5\,\mu \text{K}$, ten times lower than previously reported. $\Lambda$-enhanced cooling has also enabled us to produce optically trapped samples that are ten times higher in number and density, and forty times higher in phase space density than previously reported~\cite{Anderegg2018ODTcooling}.

\begin{figure}
\includegraphics[width=\columnwidth]{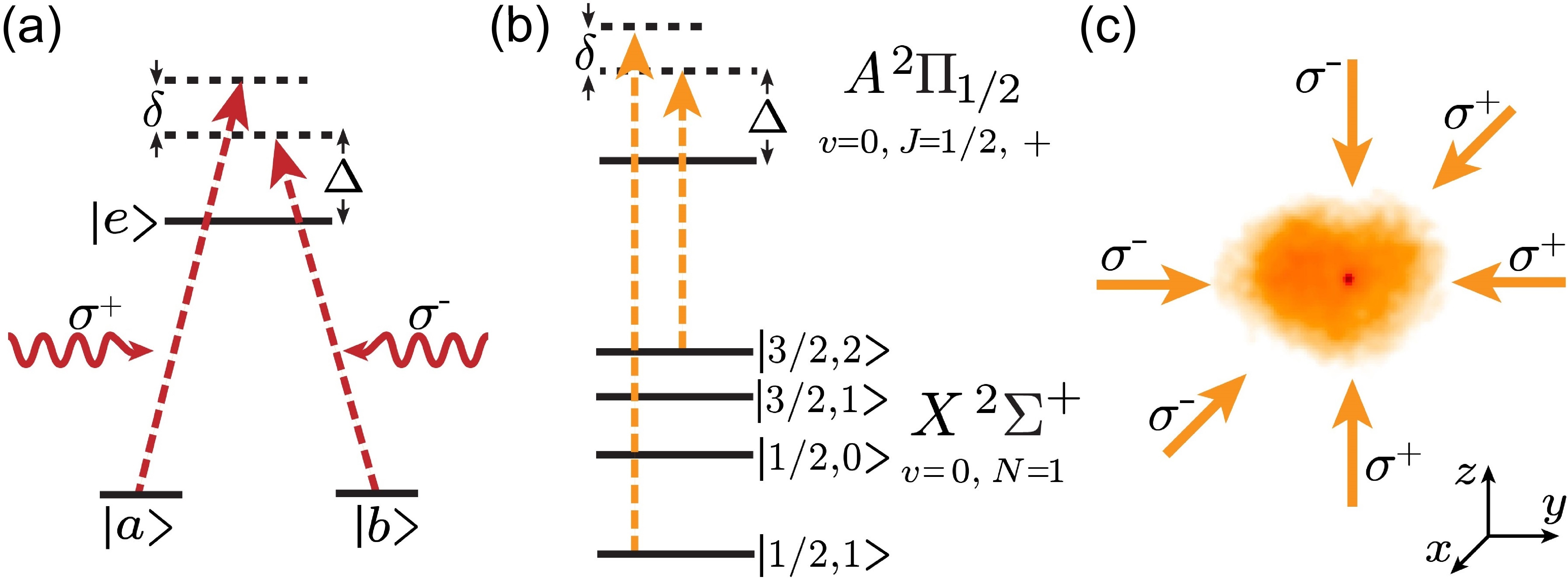}
\caption{(a) 3-level system exhibiting velocity-dependent dark states. Two ground states $|a\ket$ and $|b\ket$ are addressed separately by two counter-propagating laser beams. (b) Specific scheme for $\Lambda$-cooling of CaF. The cooling light consists of two components addressing the $|J=1/2, F=1\ket$ and $|3/2,2\ket$ hyperfine manifolds. The single-photon detuning for $|1/2,1\ket$ and two-photon detuning are denoted by $\Delta$ and $\delta$ respectively. (c) Schematic of $\Lambda$-cooling beams, overlaid with a fluorescent image of CaF molecules following $50\,\text{ms}$ of $\Lambda$-cooling. Molecules in the ODT appear as a bright spot surrounded by a much larger cloud of untrapped molecules. ($\Lambda$-imaging of trapped molecules shown in Fig.~\ref{Fig4}(c).)}
\label{Fig1}
\end{figure}

As shown recently~\cite{Truppe2017subdoppler, Anderegg2018ODTcooling}, sub-Doppler laser cooling of molecules can be achieved using gray molasses cooling, which relies on a Sisyphus cooling mechanism that appears at laser detunings to the blue of a $J\rightarrow J' (J'\leq J)$  transition~\cite{grynberg94,Sievers2015,Devlin2016}. In alkali atoms, gray molasses cooling can further be enhanced via a second mechanism that relies on velocity-dependent dark states created through two-photon resonances, a technique known as $\Lambda$-enhanced gray molasses~\cite{Grier2013D1cooling}. This second mechanism, known as velocity-selective coherent population trapping (VSCPT), has been used in atoms to reach temperatures below a single photon recoil. VSCPT cooling can be described qualitatively by a 3-level system with two ground states $|a\ket$ and $|b\ket$ addressed separately by two counter-propagating laser beams with two-photon detuning $\delta$ (Fig.~\ref{Fig1}(a)). On two-photon resonance ($\delta=0$), a dark state $\frac{1}{\sqrt{2}}(|a\ket -|b\ket)$ that is decoupled from the laser light arises for a particle at rest. A particle moving at velocity $v$ experiences a Doppler shift of the two-photon resonance of $2kv$, where $k$ is the wavevector of the light, which couples dark states to bright states. After scattering multiple photons, particles accumulate in low-velocity states, since these are longer lived than high-velocity states~\cite{Aspect1988VSCPT,Lawall19942D,Lawall19953D}. In $\Lambda$-enhanced cooling, this mechanism is further helped by standard gray molasses cooling, which can operate outside the velocity range where VSCPT is effective. In alkali atoms, $\Lambda$-enhanced cooling typically cools to temperatures of a few photon recoils, much lower than possible with gray molassses cooling alone~\cite{Grier2013D1cooling,Burchianti2014,Sievers2015,Colzi2016subdopplerNa}. 

Implementing $\Lambda$-cooling in molecules is more challenging than in alkali atoms because of more complex internal structure. For example, in CaF, the relevant states for laser cooling are comprised of four ground state hyperfine manifolds spaced by only a few excited state linewidths (Fig.~\ref{Fig1}(b)). In contrast, alkali atoms have only two ground state hyperfine manifolds that are split by 10s to 100s of linewidths. Despite these molecular complications, we have identified a simple scheme for $\Lambda$-cooling in CaF.

\begin{figure}
\includegraphics[width=\columnwidth]{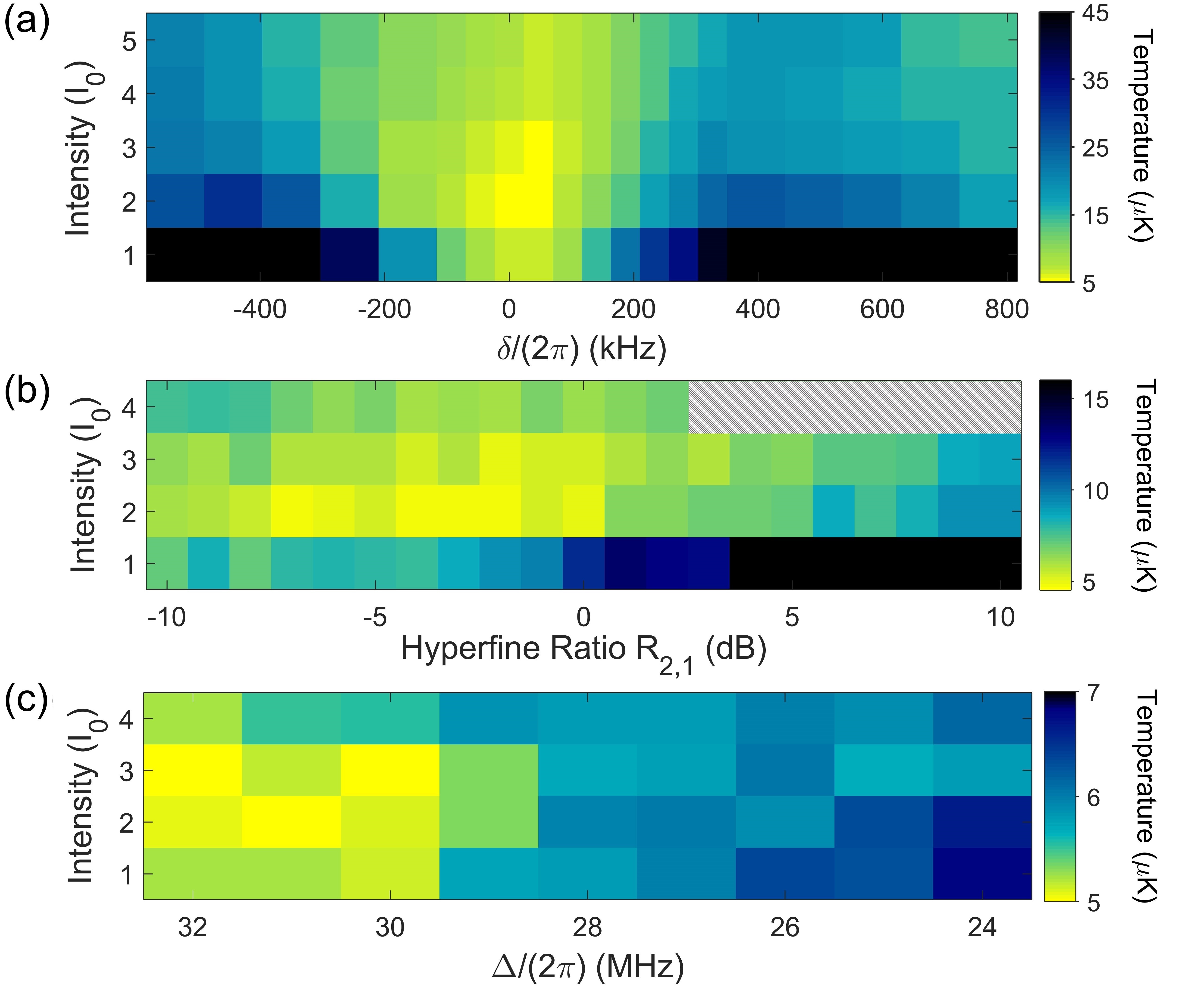}
\caption{ (a) Temperature versus intensity $I$ and two-photon detuning $\delta$ at fixed single-photon detuning ($\Delta=2.9\,\Gamma$) and hyperfine ratio ($R_{\text{2,1}}=0.92$). The hyperfine spacing between $|3/2,2\ket$ and $|1/2,1\ket$ is $h \times 73.160\,\text{MHz}$~\cite{Childs1981CaFhf}. 
(b) Temperature versus intensity $I$ and hyperfine ratio $R_{\text{2,1}}$ at fixed single-photon detuning ($\Delta=3.4\,\Gamma$). (c) Temperature versus single photon detuning $\Delta$ and intensity $I$ with $R_{\text{2,1}}=0.92$. For all plots, $I_0= 6.8\,\text{mW/cm}^2$.
}
\label{Fig2}
\end{figure}

The starting point of our experiment is a radiofrequency MOT of CaF loaded from a cryogenic buffer gas beam~\cite{anderegg17, Anderegg2018ODTcooling}. The MOT operates on the X$^2 \Sigma^+(N=1) \rightarrow$ A$^2\Pi_{1/2}(J^\prime=1/2)$ transition and consists of three retro-reflected beams, along with lasers to repump the $v=1,2,3$ vibrational levels. The MOT beams are also used for $\Lambda$-enhanced cooling. After MOT loading, we switch off the MOT beams and the magnetic gradient in 200\,$\mu$s, while simultaneously detuning the laser to $\Delta \approx +4\, \Gamma$, where $\Gamma = 2\pi \times 8.3\,\text{MHz}$ is the excited linewidth~\cite{Wall2008Astate}. The MOT beams, with polarization switching off, are then switched back on, but only with frequency components nominally addressing the $|J=3/2,F=2\ket$ and $|1/2, 1\ket$ hyperfine manifolds (Fig.~\ref{Fig1}(b)). Although only two hyperfine frequency components remain, at this detuning ($\Delta \approx 4\,\Gamma$), the $|3/2,2\ket$ component is nearly resonant with the $|3/2, 1\ket$ manifold. As a result, the light addressing $|3/2, 2\ket$ is blue-detuned by $\sim 1\,\Gamma$ with respect to the $|3/2, 1\ket$ resonance, and provides a Sisyphus cooling force. In addition, since the $|3/2,1\ket$ hyperfine manifold is nearly resonantly addressed, molecules spend a small fraction of time in these states. Thus, to a first approximation, one can ignore the effects of the $|3/2,1\ket$ manifold. For the $|1/2,0\ket$ manifold, which is also not directly addressed, the two remaining frequency components are detuned by $16\,\Gamma$ and $-6\,\Gamma$ respectively. Despite possible Sisyphus heating from the latter, this should have a small effect since the $|1/2,0\ket$ manifold consists of only one state.

Limiting to two the number of hyperfine frequency components significantly reduces the parameter space that one must search for $\Lambda$-enhanced cooling. It also allows one to gain intuition from experiments with alkali atoms. We first vary the two-photon detuning $\delta$ and the total light intensity $I$. As shown in Fig.~\ref{Fig2}(a), for all intensities used, we observe a clear temperature minimum when tuned near the two-photon resonance at $\delta=0$, with accompanying heating features when detuned. Both the heating and cooling features become more pronounced at low intensities, which can be qualitatively explained by a 3-level model. Away from resonance, the VSCPT dark states that are formed are at a finite velocity given by $\delta/(2k)$. Molecules accumulate in these longer-lived states at higher velocities, resulting in a higher average kinetic energy.
We also observe that the width of the cooling feature increases at high intensities, which is typical of VSCPT, where higher intensities increase the pumping rate into dark states. In a 3-level model (Fig.~\ref{Fig1}(a)), the bright state admixture scales as $(\delta/\Omega)^2$, $\Omega$ being the single-photon Rabi frequency. Features that vary with $\delta$ are thus expected to increase with $\Omega^2$, which is proportional to the intensity, in agreement with our observations.

Previous demonstrations of $\Lambda$-enhanced cooling of alkali atoms have reported optimal cooling when the ratio of the intensities of the hyperfine components is large~\cite{Grier2013D1cooling, Colzi2016subdopplerNa}. Which of the hyperfine components is stronger, however, was not found to be crucial~\cite{Sievers2015}. We thus explore how the ratio between the two frequency components affects temperature, which can be qualitatively different in molecules because of additional hyperfine manifolds. In contrast to observations in alkali atoms, at all intensities used, we observe a strong asymmetry with respect to the ratio of $|3/2,2\ket$ light to $|1/2,1\ket$ light, $R_{\text{2,1}}$ (Fig.~\ref{Fig2}(b)). Optimal cooling occurs when $R_{\text{2,1}}$ is between 0.2 and 1.0, at a total intensity of $I\approx 14\,\text{mW/cm}^2$. Cooling is much reduced when the $|3/2,2\ket$ power fraction is high. One possible explanation is that while the $|1/2,1\ket$ component is blue-detuned relative to all ground hyperfine states and always provides Sisyphus cooling, the $|3/2,2\ket$ component is red-detuned relative to the $J=1/2$ states and can cause Sisyphus heating. 

After optimization of the temperature with respect to the single photon detuning and the total intensity, we are able cool the molecules to $5.0(5)\mu\,\text{K}$, 8 times colder than previously reported for gray molasses cooling alone~\cite{Truppe2017subdoppler,Anderegg2018ODTcooling}. Optimal cooling is achieved at $\Delta=3.9\,\Gamma$, with a total intensity of $I=14\,\text{mW/cm}^2$, hyperfine ratio of $R_{2,1}=0.92$, and an optimal two-photon detuning of $\delta_{\text{opt,fs}}=0$. With the measured free-space density of $1.4(3)\times 10^{7}\, \text{cm}^{-3}$, the corresponding phase space density is $1.4(4)\times 10^{-8}$, 20 times higher than previously reported in free-space~\cite{Anderegg2018ODTcooling}. 

The low temperature we achieved with $\Lambda$-enhanced cooling suggested that it could be used as an imaging technique for optically trapped molecules \textemdash one can collect spontaneously scattered photons while continuously cooling. Success of this approach depends on the efficacy of in-trap cooling, which is not a given, as differential Stark shifts between ground hyperfine states could destroy coherences needed for both Sisyphus cooling and VSCPT-like dark states. 

We show here that although Stark shifts do affect $\Lambda$-cooling in an optical trap, it still remains effective. To trap molecules in our experiment, we use an optical dipole trap (ODT) formed by linearly-polarized single-frequency $1064\,\text{nm}$ light focused to a Gaussian beam waist of $29\,\mu\text{m}$. The trap light is retro-reflected with orthogonal polarization to ensure no lattice is formed. At the trapping wavelength, the differential Stark shifts are as large as $\sim 20\%$ of the trap depth. 

\begin{figure}
\includegraphics[width=\columnwidth]{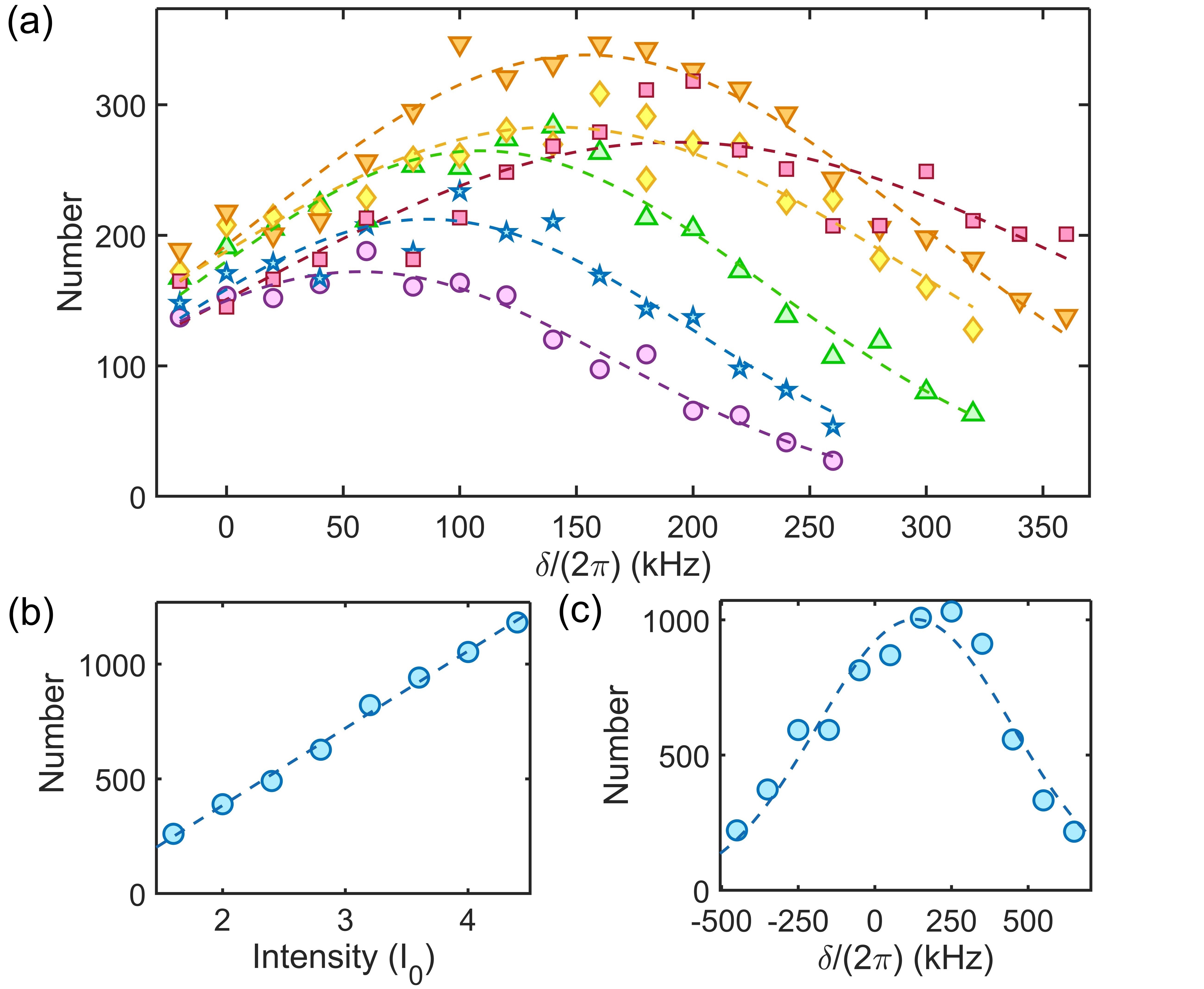}
\caption{(a) Number of molecules transferred into the ODT versus two-photon detuning $\delta$ at various trap depths $V$. With the exception of $\delta$, $\Lambda$-cooling parameters are set to the free-space optimum ($\Delta=3.9\,\Gamma$, $I=14\,\text{mW/cm}^2$, $R_{2,1}=0.92$). Transferred number for different depths are shown in purple circles ($30\,\mu \text{K}$), blue stars ($50\,\mu \text{K}$), green upward triangles ($60\,\mu \text{K}$), yellow diamonds ($90\,\mu \text{K}$), orange downward triangles ($130\,\mu \text{K}$), and red squares ($160\,\mu \text{K}$). Dotted lines show fits to a skewed Gaussian curve. (b) Transferred number versus intensity $I$ ($I_0= 6.8\,\text{mW/cm}^2$) at a trap depth of $V=k_B \times 130(10)\,\mu\text{K}$. (c) Transferred number versus two-photon frequency at trap depth of $V=k_B \times 130(10)\,\mu\text{K}$ and intensity of $I=31\,\text{mW/cm}^2$. For all plots, the single-photon detuning is $\Delta=3.9\,\Gamma$ and the hyperfine ratio is $R_{\text{2,1}}=0.92$.}
\label{Fig3}
\end{figure}

Since trap loading efficiency depends on the ability to laser cool in the trap~\cite{Anderegg2018ODTcooling}, we first explore the dependence of trapped number versus two-photon detuning at different trap depths. To transfer molecules into the ODT, we switch off the MOT beams for $200\,\mu\text{s}$ while the frequencies and intensities are adjusted. We then switch on the ODT and the cooling light, which is set $\Delta=2.9\,\Gamma$ and $I=34\,\text{mW/cm}^2$. This quickly ($1/e$ time of $1\,$ms) cools the samples down to $\sim 10\,\mu\text{K}$, which significantly reduces the expansion due to finite temperature. After $1.5\,\text{ms}$, the detuning and intensity are changed to the free-space optimum ($\Delta=3.9\,\Gamma$, $I=14\,\text{mW/cm}^2$, $R_{2,1}=0.92$) and left on for $35\,\text{ms}$. The cooling light is then switched off for $50\,\text{ms}$ to allow untrapped molecules to fall away before the number of trapped molecules is measured. As shown in Fig.~\ref{Fig3}(a), as a function of trap depth, the optimal two-photon detuning for maximal trap loading, $\delta_{\text{opt,trap}}$, is shifted from $\delta_{\text{opt,fs}}$. The range in detuning over which enhanced loading occurs increases with trap depth, and becomes broader than the cooling feature in free-space (Fig.~\ref{Fig2}(a)). The dependence of $\delta_{\text{opt,trap}}$ on $V$ at low trap depths is measured to be $+7.0(8)\times10^{-2} \times (V/\hbar)$, and saturates when at $V\approx k_B\times 130\,\mu\text{K}$. The shift in $\delta_{\text{opt,trap}}$ is of the same scale as estimated differential Stark shifts between ground hyperfine states. The saturation of $\delta_{\text{opt,trap}}$ with trap depth $V$ might arise from the competition between optimal $\Lambda$-cooling in free-space and inside the trap. In deep traps, $\delta_{\text{opt,trap}}$ can be shifted beyond the free-space cooling feature. When this occurs, one expects reduced loading efficiency.

In order to optimize for both free-space and in-trap cooling, one can use higher intensities to broaden the $\Lambda$-enhanced cooling feature at the expense of minimum attained temperature (Fig.~\ref{Fig2}(a)). To test this idea, we fix the trap depth ($V= k_B\times130(10)\,\mu\text{K}$) and the two-photon frequency ($\delta_{\text{opt,trap}}$) for optimal loading and vary the intensity $I$, and single-photon detuning $\Delta$. As shown in Fig.~\ref{Fig3}(b), the number of loaded molecules increases with intensity, consistent with the idea that increased intensity reduces the sensitivity to $\delta$. which also varies spatially in the trap due to differential Stark shifts. We find minimal dependence of $\Delta$.

To verify that the two-photon resonance remains a key factor at high intensities, we measure the loaded number versus $\delta$ at the maximum intensity available ($I=31\,\text{mW/cm}^2$). As shown in Fig.~\ref{Fig3}(c), we observe a broad enhancement feature with a width in $\delta$ of $\sim 2\pi \times 1\,\text{MHz}$. With optimized parameters ($\Delta=3.9\,\Gamma$, $\delta=2\pi \times 90\,\text{kHz}$, $I=31\,\text{mW/cm}^2$, $V=130(10)\,\mu\text{K}$), 1300(160) molecules are transferred into the ODT with a temperature of $21(3)\mu\text{K}$, 3 times colder and 9 times higher in number than previously reported without $\Lambda$-enhanced cooling~\cite{Anderegg2018ODTcooling}. The peak trapped density of $6\times10^{8}\,\text{cm}^{-3}$ and phase space density of $8(2)\times10^{-8}$ is 8 times and 40 times higher respectively~\cite{Anderegg2018ODTcooling}. The significant improvement in ODT transfer using $\Lambda$-enhanced cooling suggests that it remains effective in the trap. 

\begin{figure}
\includegraphics[width=\columnwidth]{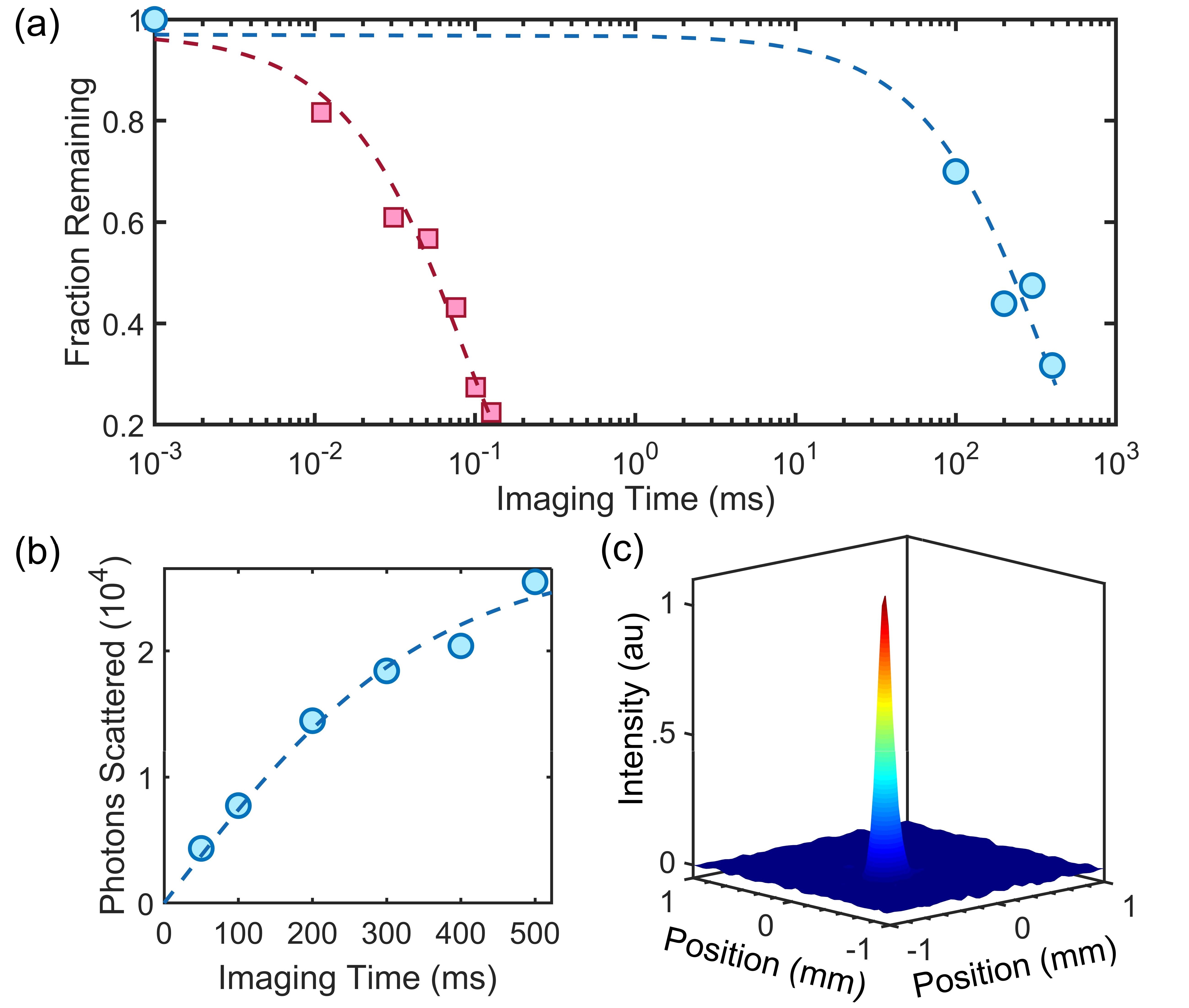}
\caption{(a) Fraction of molecules remaining versus imaging time. $\Lambda$-imaging shown in blue circles; resonant imaging shown in red squares. (b) Total number of photons scattered versus imaging time (c) In-situ $\Lambda$-imaging of trapped molecules. The exposure time is $200\,\text{ms}$, and $50$ individual images are averaged. The $\Lambda$-imaging parameters are $\Delta=3.9\,\Gamma$, $\delta=2\pi \times 90\,\text{kHz}$, $I=31\,\text{mW/cm}^2$ and $R_{2,1}=0.16$.}
\label{Fig4}
\end{figure}

To show that $\Lambda$-cooling can be a tool for non-destructive detection, we first measure the trapped number as a function of cooling time. Molecules are loaded into the ODT by first turning on $\Lambda$-cooling light for $150\,\text{ms}$, then switching it off for $50\,\text{ms}$ to allow untrapped molecules to fall away. We then switch back on the cooling light for a variable time. In order to normalize out losses due to collisions with background gas, the samples are always held for the same amount of time. We find that the imaging lifetime is sensitive to the hyperfine ratio $R_{2,1}$. At an optimal ratio of $R_{2,1}=0.16$, we measure a trap lifetime of $370(60)\,\text{ms}$ (Fig.~\ref{Fig4}(a)). By comparing the collected fluorescence with that of resonant imaging, we determine the scattering rate for $\Lambda$-cooling to be $\Gamma_{\Lambda}= 70(10)\times10^3\,\text{s}^{-1}$ for these parameters. This corresponds to the scattering of 2700(600) photons per molecule with $10\%$ loss. In contrast, with resonant imaging (scattering rate of $1.6(2)\times10^6\, \text{s}^{-1}$), the imaging lifetime is $80(5)\,\mu\text{s}$ (Fig.~\ref{Fig4}(a)), corresponding to the scattering of 13(2) photons per molecule with $10\%$ loss. $\Lambda$-imaging thus produces 200 times more photons.
We also observe that even after $150\,\text{ms}$ of $\Lambda$-imaging the temperature of the molecules remains unchanged within experimental uncertainty, staying at $20(3)\,\mu\text{K}$, $\sim 6$ times below the trap depth. In contrast, resonant fluorescent imaging applied for $60\,\mu\text{s}$ increases the temperature to $50\,\mu\text{K}$ and leads to significant losses. 

A useful metric for detection of atoms and molecules is the imaging lifetime $\tau$ normalized by the scattering rate $\Gamma_{\Lambda}$, $\xi=\tau \times \Gamma_{\Lambda}$. We find that $\Lambda$-enhanced imaging gives $\xi=2.6(6)\times10^{4}$. Two limiting mechanisms for $\xi$ are branching into vibrational states not addressed by the available repumpers ($v=1,v=2,v=3$), and mixing of $N=3$ states into the nominal $N=1$ states due the hyperfine interaction. By measuring the MOT lifetime, we determine that these mechanisms will not limit $\xi$ below $10^5$. A separate loss mechanism is diffusion in the presence of $\Lambda$-cooling, which arises when the scattering rate $\Gamma_{\Lambda}$ is much larger than the trap frequencies, which are $\omega_{x,y,z}= 2\pi \times (1.5\times10^3, 1.5\times10^3, 12)\,\text{s}^{-1}$ in our setup. This effect can be captured by a simple model where the velocity of a molecule is described by a Boltzmann distribution at a temperature of $20\, \mu\text{K}$, and randomized at the scattering rate $\Gamma_{\Lambda}$. 
A Monte-Carlo simulation taking into account trap dynamics and gravity yields a lifetime of $700(100)\text{ms}$, 2 times longer than observed. We believe that this simple model captures the dominant loss mechanism, and differences are likely explained by spatially inhomogeneous cooling. This diffusive loss could be reduced by lowering the scattering rate at the expense of longer photon collection time. 

With the imaging lifetime achieved in this work, single-shot non-destructive readout of single molecules is now possible. In future experiments, where high photon collection efficiency can be obtained using a microscope objective, we estimate that 10s of photons per molecule can be detected with imaging losses in the $1\%$ range. This projected photon number will be sufficient for high-fidelity detection of single molecules.

In conclusion, we have demonstrated non-destructive imaging of optically trapped CaF molecules using $\Lambda$-enhanced gray molasses cooling. Despite complexities in the hyperfine structure, we have identified and implemented a scheme of $\Lambda$-cooling that enables us to cool to $5\,\mu\text{K}$ in free-space. This technique has significantly improved production of optically trapped samples, allowing trapping of 1300(160) molecules at a temperature of $21(3)\,\mu\text{K}$ and a peak density of $6(2)\times 10^{8}\,\text{cm}^{-3}$. These densities are now sufficient for loading into arrays of optical tweezers, an emerging platform for quantum simulation and information processing~\cite{schlosser2002tweezerblockade,yavuz2006tweezerrabi,liu2018moltweezer,blackmore2018tweezerQI}. Despite effects from differential Stark shifts, we have found $\Lambda$-cooling to be effective in an optical trap. By collecting scattered photons during $\Lambda$-cooling, we are able to non-destructively detect trapped molecules. Compared to resonant fluorescent imaging, photon-cycling is greatly enhanced, and $200$ times more photons are emitted. Our imaging method opens the door to high-fidelity read-out of single molecules and creation of defect-free molecular arrays~\cite{lukin16array,barredo16array,Bernien201751atom}. The methods developed here are not specific to CaF, but are broadly applicable to other laser-coolable molecules (e.g. SrF, YbF, YO, YbOH, SrOH, CaOH, $\text{CaOCH}_3$), suitable for a wide variety of applications ranging from precision probes of particle physics~\cite{hinds12,Lim2017,Kozyryev2017,carr09} to ultracold chemistry~\cite{krems08, carr09, kozyryev2016lasercoolpoly}. For these applications, $\Lambda$-imaging, which increases the number of scattered photons, will also be of significant help.

\begin{acknowledgments}
This work was supported by NSF. BLA acknowledges support from NSF GRFP. LWC acknowledges support from MPHQ.
\end{acknowledgments}

\end{document}